\documentclass[twocolumn,showpacs,preprintnumbers,amsmath,amssymb]{revtex4}

\usepackage{bm}
\usepackage{graphicx,amsmath,latexsym,amssymb}
\usepackage{color}
\usepackage{graphicx}
\usepackage{amsmath}
\usepackage{amssymb}

\begin{document}

\title{Phase statistics of seismic coda waves }

\author{D. Anache-M\'enier}
\affiliation{Laboratoire de Physique et de Mod\'elisation des Milieux Condens\'es, Universit\'e Joseph Fourier/CNRS, BP 166, 38042 Grenoble, France}

\author{L. Margerin}
\affiliation{Centre Europ\'een de Recherche et d'Enseignement des G\'eosciences de l'Environnement,
Universit\'e Aix Marseille, CNRS, Aix en Provence, France}

\author{B. A. van Tiggelen}
\affiliation{Laboratoire de Physique et de Mod\'elisation des Milieux Condens\'es, Universit\'e Joseph Fourier/CNRS, BP 166, 38042 Grenoble, France}


\begin{abstract}

We report the analysis of the  statistics of the phase fluctuations in the coda of earthquakes recorded during a temporary experiment deployed at Pinyon Flats Observatory, California.  The observed distributions of the  first, second and third derivatives of the phase in the seismic coda exhibit universal power-law decays whose  exponents agree accurately with circular Gaussian statistics. The correlation function  of the spatial phase derivative is measured and used to estimate the mean free path of Rayleigh waves.

% We report the analysis of the  statistics of the phase fluctuations in the coda of earthquakes recorded during a temporary experiment
% deployed at Pinyon Flats Observatory (PFO), California. The experimental  distribution of the phase and its spatial derivatives
%  are analyzed in the light of the theory  of  Gaussian random fields. The practical measurement of the phase is discussed
% in details and the main pitfalls are underlined. Since the phase itself 
%  does not contain information on the medium heterogeneity,  we study the probability
% distribution of its first, second and third spatial derivatives. For large values, these quantities
%  obey universal power-law decays whose exponents are remarkably well predicted by circular Gaussian statistics. 
% For small values, the statistics of the derivatives are flat. The details of the transition between the plateau  and the power-law
% behavior are governed by the wavelength. The correlation function  of the first phase
% derivative along the array shows a simple algebro-exponential decay with a length scale equal to the mean free path of the waves.
% Due to the limited aperture of the array and the limited amount of data, only loose bounds are provided in this study.
% Our work suggests a new method to estimate the degree of heterogeneity of the crust.

\end{abstract}

\maketitle

%\section{Introduction}
%A seismic phase refers to a well identified arrival in a seismogram
%which can be associated to a ray trajectory inside the Earth. In the
%framework of ray theory,  the phase of the signal can be defined as $\omega (t-t_0) + \delta\phi$
%  where $\omega$ is the dominant frequency of the signal and $t$, $t_0$ denote the arrival
% and origin time, respectively;   $\delta \phi$ represents phase shifts due to diffraction effects.
 In the short-period band ($> 1$ Hz) , ballistic arrivals of seismic waves
are often masked by scattered  waves due to small-scale heterogeneities in the lithosphere. The scattered elastic waves form
the pronounced tail of  seismograms  known as the seismic coda \citep{aki69,aki75}. Even when scattering is prominent,
it is still possible to define the phase of the seismic record by introducing the complex analytic signal
$\psi(t,\mathbf{r}) = A(t,\mathbf{r})e^{\mathrm{i} \phi(t,\mathbf{r})} $, with $A$  the amplitude and $\phi$ the  phase. 
%  \citet{campillo06} has summarized recent theoretical and experimental developments
% which put forward the role of the phase in seismic scattering.
%Although, many studies have focused on the modeling of the mean field intensity 
%$I(t)= \langle A(t)^2 \rangle$ \citep[see][for review]{sato98},
In the past, many studies have focused on the modeling of the mean field intensity $I(t)= \langle A(t)^2 \rangle$
 \citep[see][for review]{sato98}. The goal of the present paper is to study the statistics of the   phase field in the coda. 
In the coda, the measured displacements result from the superposition of  many partial waves which have propagated along different
 paths between the source and the receiver. Each path consists of a  sequence of scattering events that affect  the phase 
of the corresponding partial wave in a random way.  For narrow-band signals, the phase field can therefore be written as
 $\phi(t,\mathbf{r}) = \omega t + \delta \phi(t,\mathbf{r})$, where $\omega$ is the central frequency, and $\delta \phi$ denotes
the random fluctuations. The trivial cyclic phase $\omega t$ cancels when a spatial phase difference is considered between two neighbouring points.
Spatially resolved measurements are facilitated by dense arrays of seismometers that have been set up occasionally. 
We note that the phase of coda waves has not been given much attention so far. 
The advantage of phase is that it is not affected by the earthquake magnitude, and that it contains pure information on scattering, not blurred by  absorption effects.
For the statistical analysis of amplitude and phase fluctuations
of direct arrivals, we refer the reader to e.g. \citet{zheng05}.

\section{Phase distributions}

\begin{figure}
\noindent\includegraphics[width=18pc]{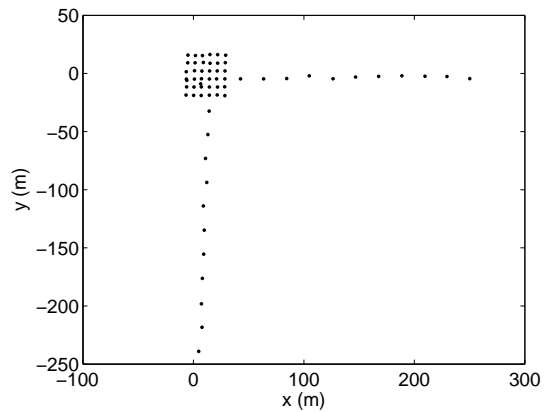}
\caption{Geometry of the seismic array}\label{rezo}
\end{figure}

We study data sets from a temporary experiment deployed at Pinyon Flat Observatory (PFO), California, in 1990 by an IRIS program. This site exhibits a high level of regional seismic activity. The array (see Fig. \ref{rezo}) contained 58 3-components L-22 sensors (2Hz) and was configured as a grid and two orthogonal arms with sensor spacings of 7 meters within the grid and 21 meters on the arms \citep{DataReport}. We selected 8 earthquakes of magnitude greater than 2 with good signal to noise ratio in the coda. Typically, epicentral distances are less
than 110 km and the coda lasts more than 30 seconds after the direct arrivals. 

To perform the statistical analysis, we filtered the signal in a narrow frequency band centered around 7 Hz ($\pm 5 \%$) and selected a 15 s time window starting around 5 seconds after the direct arrivals. In this time window, the signal is believed to be dominated by multiple scattering and is highly coherent along the array \citep{margerin08gji}. We evaluate the Hilbert transform of the vertical displacement which yields the imaginary part of the complex analytic signal $\psi(t,\mathbf{r}) = A(t,\mathbf{r})e^{i \phi(t,\mathbf{r})} $.
From the complex field, two definitions of the phase can be given: (1) The  wrapped phase $\phi$   is defined  as the argument of 
the complex field $\psi$ in the range $\left(-\pi: \pi\right]$. 
(2) The unwrapped phase $\phi_u$ is obtained by correcting for the $2 \pi$ 
jumps -- occurring when $\phi$ goes through the value $\pm\pi$ - to obtain a continuous function with values in 
$\mathbb{R}$.
% with range $\left] -\infty : + \infty \right[ $.  
The $\phi$ distribution is flat \citep{SOL}. However, more information can be extracted by considering higher-order statistics of 
the phase. For this purpose we consider the spatial derivative of the phase, which can be estimated in two different ways.

(1) The first measurement relies on the difference of the wrapped phases $\Delta \phi$ between two  seismometers separate by a distance $\delta$. Applying the simple finite difference formula $\phi' \approx \Delta \phi / \delta$ an estimate of the spatial derivative is obtained. Note that the phase difference $\Delta \phi $ takes  values between $-2\pi$ and $+2\pi$ which does not allow a precise estimate for the distribution  of the derivative for values roughly larger than $\pi/\delta$. Beyond this value our measurements will be dominated by finite difference artifacts and the distribution is biased by the $2\pi$ jumps occurring within the distance $\delta$.

\begin{figure}
\noindent\includegraphics[width=20pc]{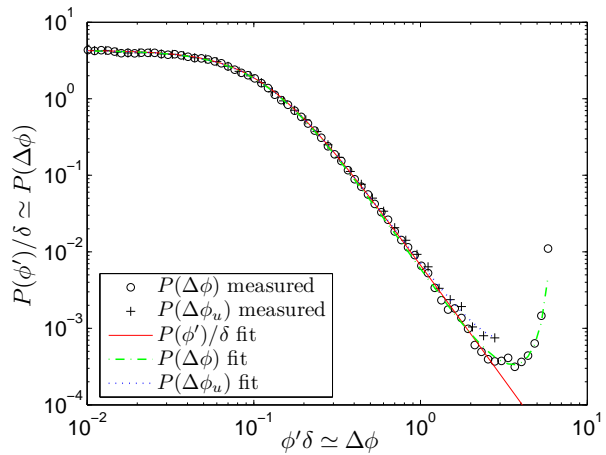}
\caption{Distribution of the first derivative of the phase normalized by the inter-station distance $\delta$ and measured using finite difference formula. ($\circ$):  wrapped phase;  ($+$): unwrapped phase.  The  colour lines represent the fits with Gaussian theory. Green:  wrapped phase; blue: unwrapped phase; red : phase derivative. 
The fitting parameters are $Q=2.774\ 10^{-4}\ m^{-2}$ and $g(\delta)=0.993204$.}\label{pphiprime}
\end{figure}

\begin{figure}
\noindent\includegraphics[width=20pc]{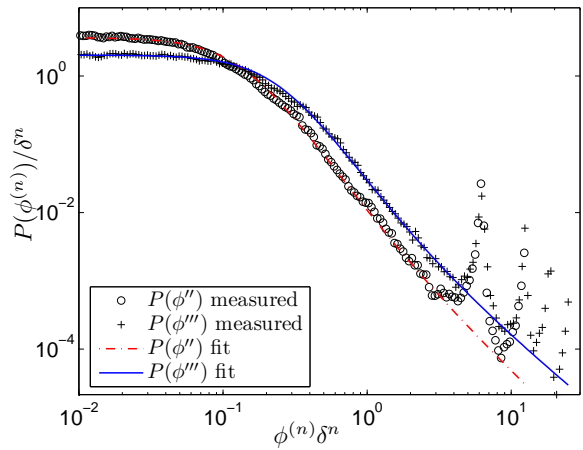}
\caption{Comparison of the Gaussian theory and observations for the phase derivatives distribution $P(\phi^{(n)})$ where $n=1$ or $2$ denotes the $n^{th}$ derivative of $\phi$ with respect to the spatial coordinate. From the fit we find: $Q=2.774\ 10^{-4}\ m^{-2}$, $R=2.75\ 10^{-2}\ m^{-2}$ and $S=4.0\ 10^{-2}\ m^{-2}$.}\label{pphideriv}
\end{figure}

(2) The second method uses the difference of the phases $ \phi_u$ {\emph spatially} unwrapped at each time step. This yields another estimate of the derivative: $\phi' \approx \Delta \phi_u / \delta$ which is expected to suppress finite difference artifact.  In practice it is impossible to discriminate a rare but physical large phase jump from a small fluctuation that causes a $2\pi$ jump just within the range $\delta$. The only possibility along 1-D arrays is to  impose that the largest admissible phase difference between two stations be smaller than $\pi$. Hence this $\phi'$ estimate takes values in $(-\pi/\delta:\pi/\delta]$ and is biased close to $\pi/\delta$ by the unwrapping processing errors.
In the limit $\delta \to 0$, the two definitions ought to be equivalent because
the probability of phase jumps between the two stations tends to 0.
% Passing to the limit requires a precise synchronization of acquisition systems. Typically, the  sampling rate should be two orders of magnitude larger than the central frequency of the waves. 
By averaging over the 8 seismic records, the lag-time in the coda, the east-west and north-south orientations, and the sensor positions within the array's grid at fixed $\delta = 7$m, we calculate the two resulting phase derivative distributions which are shown to be non-uniform in Figure \ref{pphiprime}. It is also instructive to consider the second (third) derivatives of the phase which are governed by the 3 (4)-point statistics which are plotted in Figure \ref{pphideriv}. Higher-order derivatives are obtained by applying standard finite difference formulas to the wrapped phase (this choice is explained in the next section). Since the three first derivative distributions are even functions, we only represent the positive values. They have all similar properties. For small values of the random variables, the distributions are nearly flat. For larger values, the distributions are governed by a power-law decay except for some peaks which stem from the finite distance between the seismic stations. In the following section, we will demonstrate that the transition between the two behaviours is governed by the wavelength and that the power-law decay is a very accurate signature of the Gaussian nature of the vertical displacements.
% In the section we derive the theory for a gaussian field, we show good agreement between theory and experiment for this theory and we discuss the information retreivable from the fit.

Seismic coda is believed to be composed of multiply scattered waves. Upon scattering, the many partial waves - associated with different paths in the medium - would achieve random
and independent phase shifts. The Gaussian nature would then follow from the central limit theorem. We will thus assert that the coda waves obey circular Gaussian statistics (CGS) according to which the joint probability of N complex field displacements $\psi_i$, recorded at positions $r_i$, is written as

\begin{equation} \label{eq:gauss}
P(\psi_1\cdots \psi_N)=\frac{1}{\pi^N\det
\textbf{C}}\exp\left[ - \sum_{i,j}^N \psi_i^*
\textbf{C}_{ij}^{-1} \psi_j  \right],
\end{equation}
\noindent where $\textbf{C}_{ij}=\langle \psi_i^{}
\psi_j^*\rangle $ is the covariance matrix \citep{goodman}.  It
is convenient to use normalized fields so that
$\textbf{C}_{ii}=1$. Then, the off-diagonal elements
are equal to the field correlation function $\textbf{C}_{ij}=C(\mid r_i-r_j\mid)$. 
From the joint distribution of two fields, the probability distribution of
the phase difference at two points located $\delta $ apart can be obtained by integrating over the  amplitudes \citep{PRLphase}:
\begin{equation}
\label{pdeltaphi}
P(\Delta \phi)=\tilde{P}\left[\frac{1-g^2}{1-F^2}\right]\left[1+\frac{ F\cos^{-1}\left( -F \right)}{\sqrt{1- F^2}} \right] 
\end{equation}
where $F=g\cos(\Delta\phi)$ and $g= \left<\psi(r-\delta/2) \psi*(r+\delta/2) \right>$; $\tilde{P}=(2\pi-|\Delta\phi|)/4\pi^2$ if $\Delta\phi$ is the difference of the wrapped phase and  $\tilde{P}=1/{2\pi}$ if $\Delta\phi$ is the difference of the unwrapped phase $\Delta\phi_u$. In the limit $ \delta \to \infty$ of totally uncorrelated fields $P(\Delta\phi)=\tilde{P}$. In the limit $\delta \to 0$, we get the following formula for the phase derivative \cite{Genack2a}:
\begin{equation}
\label{pphi'}
P(\phi') = \frac{1}{2}\frac{Q}{\left[Q+\phi'^2\right]^{3/2}}.
\end{equation}
where  $Q=-C''(0)= 2(1-g)/\delta^2$ for $\delta\to 0$.
Figure \ref{pphiprime} demonstrates that the  phase difference between two nearby seismometers follows the prediction of Gaussian circular statistics with  excellent accuracy: we observe a good agreement between the theoretical distribution of $\phi'$ (Eq. \ref{pphi'}) and the measurements over 3 orders of magnitude in probability and 2 orders of magnitude in derivative. A clear discrepancy occurs for large values - typically when $\phi'\approx  \pi/\delta$ - which can be perfectly explained by the finite distance between the seismic stations. This is demonstrated in the same plot which shows excellent agreement between formula (\ref{pdeltaphi}) and the measured finite-difference statistics of the phase.
 We observe that the formula for the derivative (\ref{pphi'}) agrees with the observations on a larger range when the estimate of $\phi'$ is based on the wrapped phase difference. As a consequence, we prefer this method to find the higher derivatives. 

%Note that estimating the second and third derivatives using the unwrapped phase hampers the observation of the  asymptotic power-law behaviour.

In the frequency band of interest, there is experimental evidence that the vertical component of the coda is dominated by scattered Rayleigh
surface waves \citep{margerin08gji}. As a consequence, we expect the correlation function of the field to be given by \citep{aki57}

\begin{equation}
  \label{eq:fieldcorr}
C(r)=\langle \psi(0)\psi^*(r)\rangle=\mathrm{J_0}(kr)\exp(-r/2\ell),
  \end{equation}
which agrees well with observations \citep{SOL}. Equation \ref{eq:fieldcorr} contains two  length scales: the wavelength $2 \pi/ k $ and the scattering mean free path $\ell$. We emphasize that formula (\ref{eq:fieldcorr}) is valid even when the medium is slightly inelastic.
Weak absorption does not enter the correlation function obtained from the long-time-tail of a diffuse wavefield \citep{EPLphase}. 
The form of the correlation function (\ref{eq:fieldcorr}) implies that $Q\simeq k^2/2$, if $k\ell\gg 1$. Using the parameter $Q$ obtained by fitting the data with equation (\ref{pphi'}), we infer a dominant wavelength $\lambda$ of the order of $267$ m which agrees with the one predicted on the basis of the vertical profile of the elastic constants below the PFO array \citep{vernon98}.
$Q$ offers an accurate way of estimating the wavelength, alternative to SPAC measurements \citep{aki57}.
The use of a narrow band signal is crucial because the parameters $g$ and $Q$ strongly depend on frequency.

From the joint Gaussian distribution of 3 and 4 fields, we can derive analytically the joint probability functions
$P(\phi', \phi''),$ $P(\phi', \phi'',\phi''')$ , featuring two new constants $R$ and $S$ \citep{PRLphase}. From these formulas, the marginal distributions
$P(\phi'')$, $P(\phi''')$ can be evaluated numerically.
The probability distributions of the first, second, and
third derivatives of the phase  exhibit an asymptotic power-law decay with exponents $-3$, $-2$, $-5/3$, respectively.
These universal exponents (i.e. independent of the medium properties)  can be obtained analytically 
and provide a sensitive fit-independent test of the Gaussian nature of the seismic coda. 
% The distributions of the first 3 derivatives of the phase depend on three  parameters denoted
% by $Q,$ $R$ and $S$ which are related to the  coefficients of the Taylor expansion of the
% field correlation function $C(r)$ \citep{PRLphase}:
% \begin{linenomath*}
% \begin{equation}
%   \label{eq:qrs}
% \left\{
% \begin{array}{ccl}
% C''(0) & =  & -Q \\
% C^{(4)}(0) &= & QR+Q^2\\
% C^{(6)}(0) & = & -QRS-Q(R+Q)^2\\
% \end{array}\right.
%  \end{equation}
% \end{linenomath*}
% For our data set, we obtain $C(r)=1-1.387\ 10^{-4}r^2+3.21\ 10^{-7}r^4-7.21\ 10^{-10} r^6$.
This is illustrated in  Figure \ref{pphideriv}. The agreement leaves no doubt that coda waves are in the multiple scattering regime.

% 
% The same analysis led in a systematic way for all frequencies between 4 and 30 Hz reveals also a gaussian behaviour for frequencies in the range 4-12 Hz and 22-30 Hz. Between 13 and 21 Hz, the asymptotic power law decay doesn't follow the predicted exponent. We interpret this discrepancy by... [HELP ! plizzz :-)]

%The same analysis has been carried out  in a systematic way for all frequencies between 4 and 30 Hz, and reveals  a similar Gaussian behaviour for frequencies in the range 4-12 Hz and 22-30 Hz. Between 13 and 21 Hz, the asymptotic power law decay does not follow the predicted exponent. This frequency range corresponds to brutal changes in the ellipticity  of the Rayleigh eigenfunctions. 
%This change depends on the local stratification and does not occur at the same frequency all over the array.
%In the 13-21 Hz band, we may therefore question the validity of the spatial average performed in our analysis .
% This may explain the failure of  Gaussian statistics  in some frequency bands.

\section{Phase Difference Correlations}
%\subsection{Theory}
\begin{figure}
% Correlation champ
\end{figure}
We have shown that the distributions of the phase derivatives provide accurate information on the short-range correlation
properties of the field. In the following we use the phase difference correlations  to put some constraints
on the degree of heterogeneity which is responsible for long-range correlations along the array.
For Gaussian statistics and for surface waves obeying Eq.~(\ref{eq:fieldcorr}),  the phase derivative 
correlation function is \citep{EPLphase}:

\begin{equation}\label{eq:dasym}
C_{\phi'}(r)  \equiv \left<\phi'(0) \phi'(r)\right> =    ({k}/{\pi r} ) \exp(-r/\ell)
\end{equation}
for $r>\lambda$. Formula (\ref{eq:dasym}) has one crucial property. Contrary to the field correlation function $C$, 
$C_{\phi'}$ does not oscillate on the wavelength scale but decreases with the mean free path as the
sole characteristic length scale. Any determination of  the mean free path based on formula  (\ref{eq:fieldcorr})
 is impossible because the exponential decay is masked by the rapid cyclic oscillations. 
 
 %We will show that  the phase difference correlations allow an estimate of   the mean free path on a scale of order $\ell/10$.

%\subsection{Data processing}

As above, we  estimate the phase derivative correlation function in a finite difference approximation using the formula $C_{\phi'}(r)\simeq \langle(\Delta\phi_u(r')\Delta\phi_u(r'')\rangle/\delta^2=C_{\Delta\phi_u}(|r'-r''|)/\delta^2$. Contrary to the probability distribution, we found out that the unwrapped phase difference offers a better estimate for the phase derivative. The wrapped phase difference correlation function is too much
 dominated by  large $2\pi$ jumps with no physical interest.

\begin{figure}
\noindent\includegraphics[width=20pc]{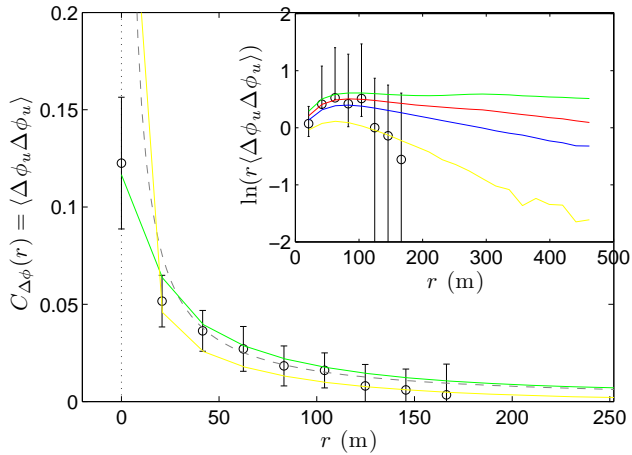}
\caption{Unwrapped phase difference correlation function. ($\circ$): experimental results. Dashed line: $1/r$ fit. Inset:  logarithm of the correlation multiplied by $r$ for data ($\circ$ with error bars)  and numerical simulations at fixed $k$ and $\ell=10$ km (green), $\ell=1$ km (red), $\ell=500$ m (red) and $\ell=200$ m (yellow). 
%The value at $r=0$ m is not represented because it is undefined.
}\label{fig:corrprime}
\end{figure}

The unwrapped phase difference correlation $C_{\Delta\phi_u}(r)$ is measured along the two orthogonal  arms with an aperture of  $252$ m  and $\delta = 21$ m interstation distance. The data are averaged over orientation, lag-time in the coda  and seismic events. The result is presented in linear scale in the main part of Figure \ref{fig:corrprime}  and shows a decay dominated by the $1/r$ factor along the arms of the array,  as predicted by  formula (\ref{eq:dasym}). This  supports our 2D picture of wave diffusion. Due to the finite difference $C_{\phi'}(r=0)$ achieves a finite value at $r=0$, which we found to be consistent with the variance
 of the unwrapped phase difference calculated from
 $\langle\Delta\phi_u^2\rangle=\int_{-\pi}^{\pi}d\Delta\phi_u\,(\Delta\phi_u)^2P(\Delta\phi_u)$ and 
$g(\delta ) =0.98$. The parameter $g(\delta)$  has been determined independently by fitting the observed distribution of $P(\Delta\phi_u)$
with formula (\ref{pdeltaphi}) for $\delta=21$ m.

Different reasons exist why it is  more difficult to measure  the  correlation function of the phase derivative  than e.g. the probability
distributions. First, 4th-order statistics  require   more averaging
to suppress unwanted  fluctuations in the data. 
Secondly, the interstation distance  $\delta=21$ m 
along the arms reduces  the correlation between the fields at two nearby stations significantly, which favours systematic errors in the derivative. 
Finally, due to frequent breakdowns of the sensors located near the ends of the arms,  the data  could  not be averaged over all sensor 
positions and all seismic events. As a consequence the correlations for distances   $r>180$ m had to be excluded.

%\subsection{Numerical simulations vs. experimental results}

Since formula \ref{eq:dasym} is valid only in the limit $\delta \to 0$, we have 
evaluated the impact of the finite inter-station distance inherent to our experimental
set-up. To this end, we have simulated N correlated Gaussian random field displacements on a virtual array  with the PFO geometry.
The results for different values of  $\ell$ at fixed $k$ are shown in Figure \ref{fig:corrprime} together with the experimental results. The simulated function $\log ( r \times C_{\Delta\phi_u})$ exhibits a linear decay with a slope $-1/\ell$ as was seen for $\delta \to 0$ in Eq.~(\ref{eq:dasym}) (see inset in Figure \ref{fig:corrprime}). Therefore, correlations of finite-size phase difference  still offer  direct access to the mean free path of the waves in the crust.
% Note that the variance of the unwrapped phase difference (value at $r=0$) is not a good candidate to probe the mean free path
% because it also depends on the correlation length of the disorder \citep{rytov89}.  
It is quite encouraging to see (inset Figure \ref{fig:corrprime}) that the asymptotic exponential regime is  already reached 
for $r>\lambda/5$. We find that the $-1/\ell$ slope
could in principle be measured if the aperture of the network would have been a few wavelengths in size ($> 500$ m), which in general is much
smaller than the mean free path. Unfortunately, the experimental error bars of our data set are too large
 to permit accurate estimates of $\ell$ at PFO although it can roughly be bounded between 1 km and 10 km, much  smaller than the typical  path (40 km)  taken by coda waves coda during 20 s. This has to be interpreted as a rough estimate
for the mean free path of  the Rayleigh waves.\\

In conclusion, from their observed first three  spatial derivatives of phase, seismic coda waves are proved to obey Gaussian statistics  with high accuracy, with a local correlation on the scale of the wavelength. At longer scales, we demonstrate that the correlation function of the spatial derivative of phase offers a new, promising opportunity to measure directly the  mean free path $\ell$ of seismic waves. This measure is not affected by absorption neither to scattering anisotropy. As opposed to the coherent backscattering effect \citep{cone},  the proposed method does not require the sensors to be located in the near field of the source, but requires an array of seismic stations  with sub-wavelength spacing between stations
and with a total aperture of a few wavelengths. Our work highlights the phase as a useful physical object
 to study seismic coda waves.

We thank J. Page, M. Campillo, P. Roux and E. Larose for useful discussions.

\bibliography{bibgrl}

\end{document}